\documentclass[prl,aps,groupedaddress,superscriptaddress,notitlepage, twocolumn]{revtex4-1}

\usepackage[utf8x]{inputenc}
\usepackage{color}
\usepackage{bbm} 

\usepackage{nicefrac}
\usepackage{amsfonts,amsmath,amssymb,stmaryrd}
\usepackage{braket}

 \usepackage[autostyle=true]{csquotes}

\usepackage{tabularx}
\usepackage{multirow} 
\usepackage{hhline}
\usepackage{graphicx}
\usepackage{subfigure}  % use for side-by-side figures
\usepackage{bbm} 
\usepackage{hyperref}
\usepackage[pdftex]{epsfig}
\usepackage{mathrsfs}
\usepackage{verbatim}
\usepackage{centernot}
\usepackage{ulem}
\usepackage{array}
\usepackage{cancel}
\usepackage{ifthen}
\usepackage{bm}	% bold in math mode
\usepackage{siunitx}
\usepackage{float} % to use [H] in \begin{figure}[H]
\usepackage[super]{nth} % use for expressions 3rd, 10th....  
\usepackage{todonotes}
\usepackage{color,soul}
\InputIfFileExists{gitinfo-latex.inc}{}{}

\usepackage{listings}
\usepackage{color}
\usepackage[acronym,nomain]{glossaries}

\begin{document}
	
\title{An endoreversible quantum heat engine driven by atomic collisions}
\author{Quentin Bouton}
\thanks{J.N. and Q.B. contributed equally to this work.}
\affiliation{Department of Physics and Research Center OPTIMAS, Technische Universit\"at Kaiserslautern, Germany}

\author{Jens Nettersheim}
\thanks{J.N. and Q.B. contributed equally to this work.}
\affiliation{Department of Physics and Research Center OPTIMAS, Technische Universit\"at Kaiserslautern, Germany}

\author{Sabrina Burgardt}
\affiliation{Department of Physics and Research Center OPTIMAS, Technische Universit\"at Kaiserslautern, Germany}

\author{Daniel Adam}
\affiliation{Department of Physics and Research Center OPTIMAS, Technische Universit\"at Kaiserslautern, Germany}

\author{Eric Lutz}
\affiliation{Institute for Theoretical Physics I, University of Stuttgart, D-70550 Stuttgart, Germany}

\author{Artur Widera}
\email{email: widera@physik.uni-kl.de}
\affiliation{Department of Physics and Research Center OPTIMAS, Technische Universit\"at Kaiserslautern, Germany}
\date{\today}

\begin{abstract}
Quantum heat engines are subjected to quantum fluctuations related to their discrete energy spectra.
Such fluctuations question the reliable operation of quantum engines in the microscopic realm.
We here realize an endoreversible  quantum Otto cycle in the large quasi-spin states of Cesium impurities immersed in an ultracold Rubidium bath.  Endoreversible machines are internally reversible and irreversible losses only occur via thermal contact.  We employ quantum control over both machine and bath to suppress internal dissipation and regulate  the direction of heat transfer that occurs via inelastic spin-exchange collisions. We additionally use  full-counting statistics of individual atoms to monitor heat exchange between engine and bath at the level of single  quanta, and evaluate  average and variance of the power output. We optimize the performance as well as the stability of the  quantum  engine, achieving high efficiency, large power output and small power output fluctuations.

\end{abstract}

\maketitle

Most engines used in modern society are heat engines. Such machines  generate motion by converting thermal energy into mechanical work \cite{cen01}. Two central figures of merit of heat engines are efficiency, defined as the ratio of work output and heat input, and power  characterizing the work-output rate. Heat engines should ideally have high efficiency, large power output, and  be stable, i.e., exhibit small power fluctuations. However,  real thermal machines operate far from reversible conditions and their performance is thus reduced by irreversible losses \cite{and84,and11}. At the same time, microscopic motors are exposed to thermal fluctuations and, at low enough temperatures, to additional quantum fluctuations, which are associated with random transitions between discrete levels. Both fluctuation mechanisms contribute to their instability \cite{pie18,hol18}. An important issue is hence to design and optimize small heat engines in order to maximize both their performance and their stability \cite{den20}.

Nanoscopic heat engines have been  implemented recently using a single trapped ion \cite{Rossnagel_2016} and  a spin coupled to the single-ion motion \cite{Lindenfels2019,hor20}. Indications for quantum effects have been reported in a spin engine consisting of nitrogen-vacancy centers interacting   with a light field \cite{Klatzow19}, and quantum heat engine operation has been shown in nuclear magnetic resonance  \cite{Assis_2019,Peterson19} and single-ion  \cite{hor20} systems.
These thermal machines are based on  harmonic oscillators or  two-level systems, and the baths mediating heat exchange are   simulated by interaction with either laser  fields \cite{Rossnagel_2016,Lindenfels2019,hor20,Klatzow19} or radiofrequency pulses \cite{Assis_2019,Peterson19}.

We here experimentally realize a quantum Otto cycle using a large quasi-spin system in individual Cesium (Cs) atoms immersed in a natural quantum heat bath made of ultracold Rubidium (Rb) atoms.
Expansion and compression steps are implemented by varying an external magnetic field, changing the energy-level spacing of the engine and performing work \cite{kos17}. Heat exchange between system and bath occurs via inelastic endoenergetic and exoenergetic spin-exchange collisions \cite{Schmidt2019}.   
The increased number of internal engine states, compared to simple two-level systems,  allows for high energy turnover per cycle,  while their  finite number naturally limits power fluctuations due to saturation, in contrast to the unbound spectrum of harmonic oscillators.   
We employ quantum control of both the engine's quasi-spin state and the bath's spin polarization to control the direction of heat transfer between the two at the level of individual quanta of heat \cite{Schmidt2019}, independent of the kinetic thermal state of the bath. This quantum control effectively suppresses internal irreversible losses at the level of individual collisions and thus make the quantum heat engine endoreversible. 
Endoreversible machines operate internally without any dissipation, while irreversible losses only occur via the contact with the bath. They outperform fully irreversible engines and have played a central role in finite-time thermodynamics for forty years \cite{and84,and11}.  
We additionally characterize the discrete quantum heat transfer at the level of individual quanta using full-counting statistics \cite{esp09,fri18} and monitor the population dynamics of the  engine from single-atom  and time-resolved measurements of the engine's quasi-spin distribution along the cycle. %\cite{Bouton2020}. 
We employ this new system and novel techniques to evaluate and optimize the performance as well as the stability of the endoreversible quantum heat engine, achieving  high efficiency, large power output  and small power output fluctuations.

\begin{figure*}
\label{fig:IntroSketch}
\includegraphics[width=0.88\textwidth]{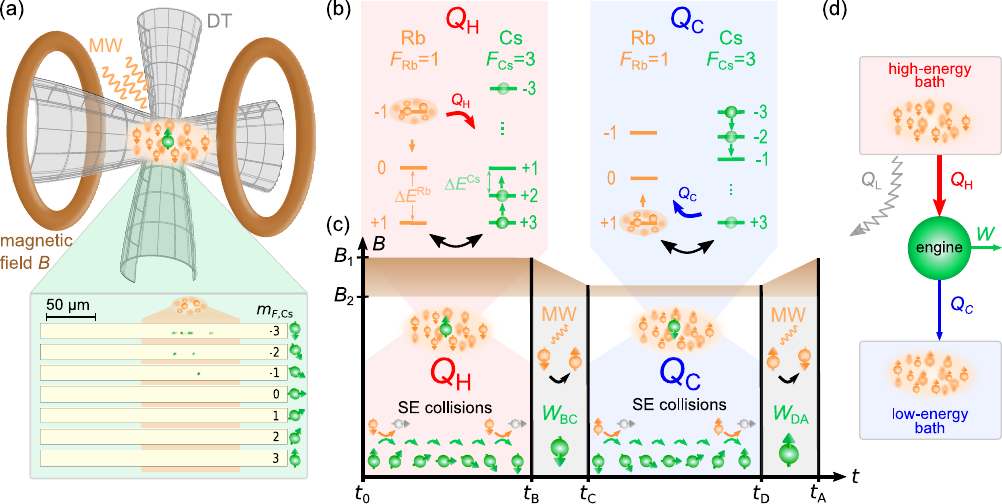}
\caption{\textbf{Operation principle of the endoreversible quantum heat engine.} (a) Individual laser-cooled Cs atoms (green) are immersed in an ultracold Rb cloud (orange); both are confined in a common optical dipole trap (DT). External magnetic fields and microwave (MW) radiation respectively implement the power strokes of the quantum heat engine and distinguish the high- from the low-energy bath. The inset shows typical $m_F$- resolved fluorescence images of single Cs atoms for $t= t_B=300\,$ms after initialization. The position of the  bath cloud is indicated in orange with a width of $4 \sigma$.
(b) The heat exchange between the Cs atom (engine) and a Rb (bath) atom occurs via  inelastic spin-exchange collisions. Spin polarization of the Rb  atoms and spin-conservation in individual collisions allow only up to six exo- or endothermal processes, corresponding to heating or cooling. 
(c) The experimental Otto cycle consists of a  heating stage, during which heat $Q_\text{H}$ is absorbed, and a power stroke induced by an adiabatic change of the magnetic field.  A microwave field then switches the bath from high to low energy. The cycle is further completed by a cooling step, during which heat $Q_\text{C}$ is released, and an additional power stroke when the magnetic field is adiabatically brought back to its initial value.
(d)  Due to the difference of atomic Land\'e factors between Cs and Rb, the quantum heat engine (green) absorbs heat $Q_\text{H}$ and releases heat $Q_\text{C}$ (to produce  work $W$), while the bath releases more energy.  The lost energy is irreversibly dissipated during an average of ten elastic collisions and is described by a heat leak $Q_\text{L}$ from the high-energy bath.}
\end{figure*}

We experimentally immerse up to ten laser-cooled Cs atoms in the $\ket{F_\text{Cs}=3,m_{F,\text{Cs}} = 3}$ state into an ultracold Rb gas of up to $10^4$ atoms in the state $\ket{F_\text{Rb}=1,m_{F,\text{Rb}}=-1}$, both species confined in a common optical dipole trap   (Fig.~1(a)) (Appendix A).
Here $F$ and $m_F$ denote the total atomic angular momentum and its projection onto the quantization axis, respectively. % \cite{coh11}. 
 The quantization axis is given by an external magnetic field of $B_1 = 346.5 \pm 0.2\,$mG or $B_2 = 31.6 \pm 0.1\,$mG. 
The Cs atoms quickly thermalize to the kinetic temperature of $T=950\pm 50\,$nK of the gas. % \cite{Bouton2020}. 
We operate the quantum heat engine in the spin-state manifold of the seven Cs hyperfine ground states $\ket{F_\text{Cs}=3,m_{F,\text{Cs}}}$, $m_{F,\text{Cs}}\in [+3, +2, \ldots, -3]$. 
The states are energetically equally spaced with Zeeman energy $E_n^\text{Cs} = n \lambda B$, with $\lambda= |g_F^\text{Cs}| \mu_\text{B}$, where $g_F^\text{Cs}=-1/4$ is the Cs Landé factor, $\mu_{B}$ Bohr's magneton and $n=3-m_{F,\text{Cs}}$ \cite{coh11}, with the zero-point of energy set to the lowest-energy state $\ket{m_{F,\text{Cs}}=3}$.

Heat between the quantum   engine and the bath is  exchanged at the microscopic level via inelastic spin-exchange collisions (Fig.~1(b)). 
Each collision changes the value of the quasi-spin of the Cs engine by $\Delta m_\text{Cs}=\mp 1 \hbar$ leading to an energy change of $\Delta E^\text{Cs} = \pm  \lambda B$ for each Cs atom, and $\Delta m_\text{Rb} = \pm 1 \hbar$ for one Rb atom corresponding to the energy change $\Delta E^\text{Rb} = \mp \kappa B$, with $\kappa= |g_F^\text{Rb}| \mu_\text{B}$, where $g_F^\text{Rb}=-1/2$ is the Rb Landé factor \cite{Schmidt2019}. 
The spin population thus directly reflects the energy exchange between engine and reservoir at the level of single energy quanta.
The direction of the heat transfer is determined by the spin polarization of the Rb bath and by angular momentum conservation during individual collisions.  
The spin polarization of the Rb atoms distinguishes a high-energy  bath for $m_\text{Rb}=-1$ from a low-energy bath for $m_\text{Rb}=+1$. Control over the internal Rb state accordingly permits to either increase or decrease the energy of the quasi-spin of the engine.
Heat exchange automatically stops after six spin-exchange collisions, because then the highest/lowest energy state has been reached. The collision transfers the colliding Rb atom to the $\ket{F_\text{Rb}=1,m_{F,\text{Rb}}=0}$ state, which forms the exhaust of the engine. 
Due to the massive imbalance between the Rb and Cs atom numbers ($N_\text{Rb}/N_\text{Cs} > 1000$), each collision occurs with a bath atom in the initial state with high probability, making the bath Markovian.

\begin{figure*}
\label{fig:QuantumEngine}
\includegraphics[width=0.9\textwidth]{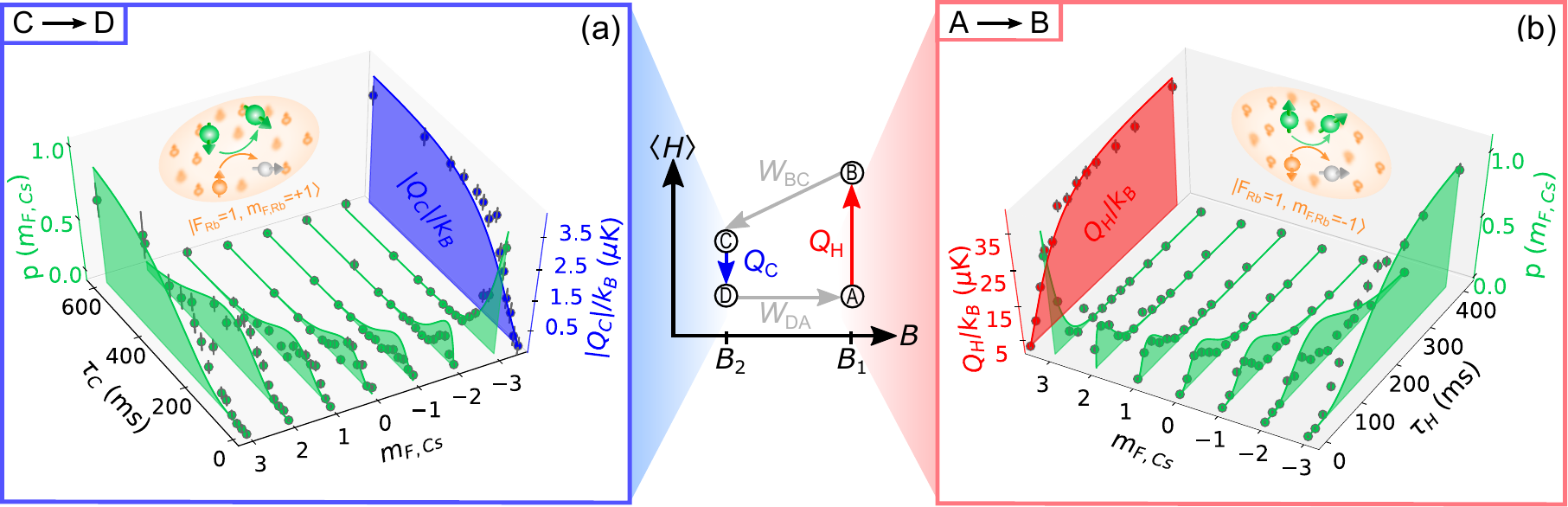}
\caption{\textbf{Full-counting statistics of heat exchange.}
During the heating (AB) and cooling (CD) steps  of the quantum Otto cycle (center), heat is exchanged with the bath. The population dynamics  of the individual engine levels  are shown in green.  The averages of heat,  $Q_\text{H}$ and $Q_\text{C}$, extracted from the full counting statistics are indicated for (a) cooling (blue) and (b)  heating (red), as a function of the respective times $\tau_\text{H}$ and $\tau_\text{C}$. Dots show the experimental data, solid lines are a prediction of a microscopic model (Appendix C). In both panels, the population dynamics shows the transition from an initially spin-polarized engine state via a state of many populated $m_F$ levels to a spin polarized state of the other extreme spin state. The inversion of an initially fully polarized population ($\ket{m_{F,\text{Cs}}=3} \leftrightarrow \ket{m_{F,\text{Cs}}=-3}$) requires some hundreds of milliseconds. 
}
   \end{figure*} 
   
The quantum Otto cycle consists of four parts: one compression and one expansion step, during which work is performed, and a heating and a cooling stage, during which heat is exchanged \cite{kos17}. 
The corresponding experimental sequence is shown in Fig.~1(c). The Cs machine is first driven by up to six spin-exchange collisions into  energetically higher states (at magnetic field $B_1$),  absorbing heat $Q_\text{H}$ in time $\tau_\text{H}=t_\text{B}$.  Work $W_\text{BC}$ is then performed by adiabatically decreasing the magnetic field to $B_2$ in $\tau= t_\text{C}-t_\text{B}=10\,$ms. This time is much longer than the inverse energy splitting $\Delta E$ of the  quasi-spin states, making the process adiabatic. It is, however, fast enough to avoid unwanted spin-exchange collisions, implying that no heat is transferred. 
The engine is subsequently brought into contact with the low-energy bath by flipping the spins of the Rb bath using microwave (MW) sweeps. The Cs engine is accordingly driven by up to six spin-exchange collisions into  energetically lower  states, releasing heat $Q_\text{C}$ in time $\tau_\text{C}=t_\text{D}-t_\text{C}$. Work $W_\text{DA}$ is further performed by adiabatically increasing the magnetic field back to $B_1$ in $\tau = t_\text{A} -t_\text{D}=10\,$ms. The Rb spins are finally flipped to their initial state with other microwave sweeps, restoring the high-energy bath.

While each single collision is coherent and thus amenable to quantum state engineering, coupling of the engine to the large number of bath modes in elastic collisions destroys the coherence between the engine's quasi-spin levels. Heat is thus associated with changes of occupation probabilities, $Q=\sum_n E_n \Delta p_n$, whereas work corresponds to changes of energy levels, $W=\sum_n p_n \Delta E_n$ \cite{kos17}. In our system, we concretely have $Q_\text{H}=\sum_{n}n \left(p_{n}^\text{B}-p_{n}^\text{A} \right)\lambda B_{1}$ for the heating process and $Q_\text{C}=\sum_{n}n \left(p_{n}^\text{D}-p_{n}^\text{C} \right) \lambda B_{2}$ for the cooling process. 
On the other hand, the respective work contributions for expansion and compression are given by $W_\text{BC} = \sum_{n}n p_{n}^\text{B}\lambda (B_{2}-B_{1})$ and $W_\text{DA} = \sum_{n}n p_{n}^\text{D} \lambda (B_{1}-B_{2})$. 
In order to evaluate these quantities,
we determine the  magnetic fields $B_1$ and $B_2$  with the help of Rb microwave spectroscopy (Appendix A). We further detect the Zeeman populations $p_n^i$ of individual Cs atoms at arbitrary times by position resolved fluorescence measurements combined with Zeeman-state selective operations \cite{Bouton2020}. 
From a series of such measurements, we can, atom by atom, construct the quasi-spin populations at any time (Fig.~2). This allows us to monitor for the first time the discrete heat exchange between engine and environment with a resolution of single quanta at each time (Fig.~2): the progressive transfer from low (high) energy states to high (low) energy states during heating (cooling) as a function of time is clearly seen (green dots). 
From the measured heat counting statistics, we compute average (blue and red dots) and variance of heat exchange (Appendix D). We will use these quantities to examine the power output of the quantum machine and its fluctuations.

\begin{figure}
\centering
\includegraphics[width=8cm]{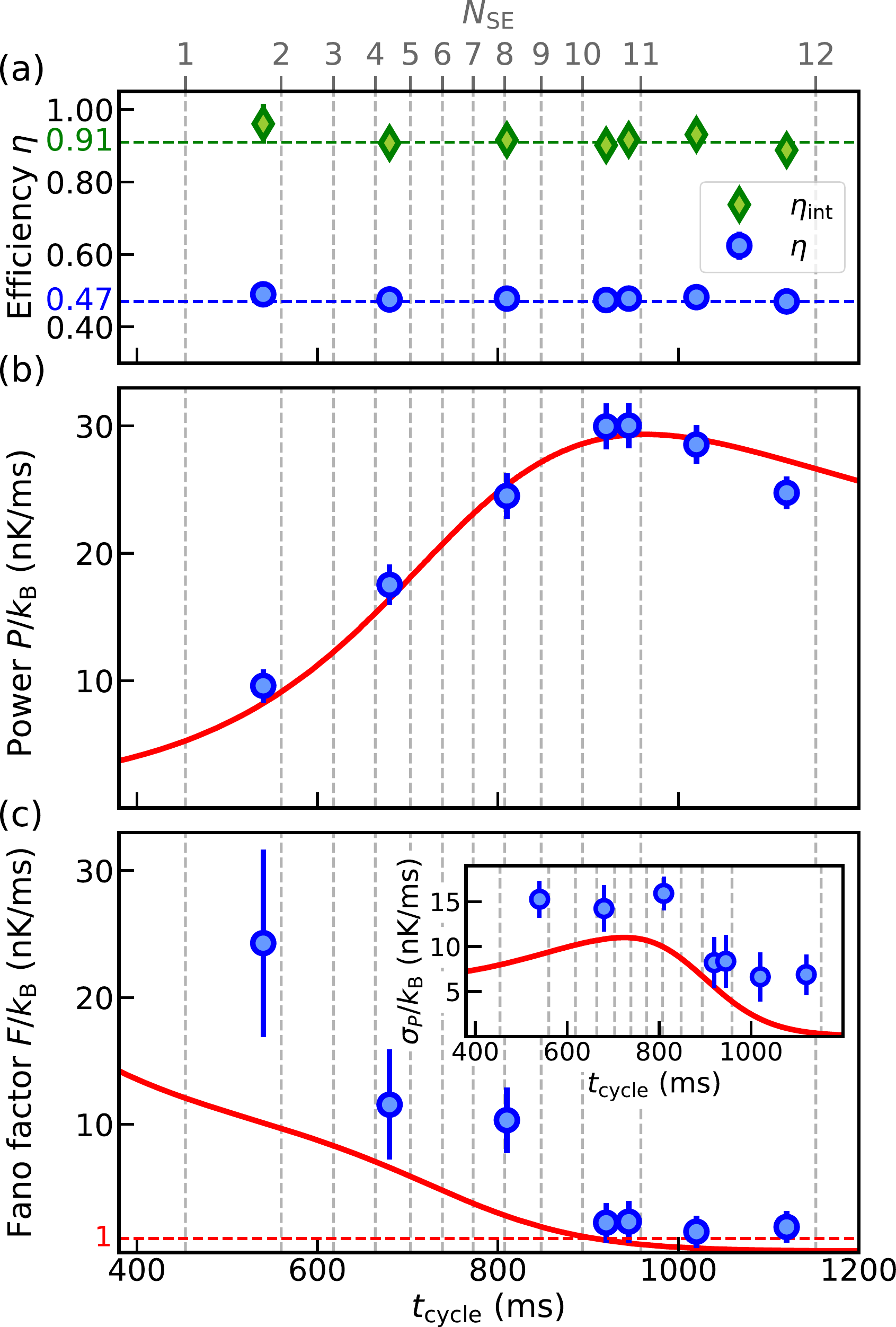}
\caption{\textbf{Performance of the quantum heat engine.}
(a) Efficiency $\eta$, Eq.~\eqref{eq:efficiency_dissip} (blue dots), and internal (dissipationless) efficiency $\eta_\text{int}$ (green diamonds) for different cycle times; dashed lines indicate the respective expected values. (b) Power output, Eq.~(\ref{eq:power}) (blue dots: experimental data, red solid line: theoretical model), with maximal value reached after almost 12 spin exchange collisions.  (c) Fano factor, Eq.~(\ref{eq:Fano_factor}), and time-resolved fluctuations $\sigma_P$ (inset).  In all cases, the dashed vertical lines (upper axis) indicate the number of spin-exchange collisions $N_\text{SE}$.
The different durations between two successive spin-exchange collisions originate from different atomic transition rates \cite{Schmidt2019}.   
}
\label{fig:Timeevolution}
\end{figure}

We first characterize the performance of the quantum Otto engine by evaluating its efficiency given by \cite{kos17},
\begin{equation}
\eta= \frac{Q_\text{H} - \vert Q_\text{C}\vert }{Q_\text{H} +  Q_\text{L}}, \label{eq:efficiency}
\end{equation}
where $Q_\text{H} - \vert Q_\text{C}\vert$ is the total work produced by the thermal machine, $Q_\text{L}$ the energy dissipated during the total heat exchange in one cycle, and $Q_\text{H}+Q_\text{L}$ the heat emitted by the high-energy bath (Fig.~1(d)). Indeed, due to the different  atomic Land\'e factors for Rb $(g_F^\text{Rb}=-1/4)$ and Cs $(g_F^\text{Cs}=-1/2)$, only half ($\gamma = g_F^\text{Cs}/g_F^\text{Rb}= 1/2$) of the energy change of a bath atom is effectively exchanged with the  heat   engine during an inelastic spin-exchange collision \cite{Bouton2020}.  
As a result, the heat emitted (absorbed) by the bath differs from the energy portions absorbed $Q_\text{H}$ (emitted $Q_\text{C}$) by the machine. We macroscopically account for  the remaining lost energy, which is irreversibly dissipated during an average of ten elastic collisions, by a heat leak  \cite{mas19} equal to $Q_\text{L}=\sum_n n \left (p_n^B - p_n^A \right) \kappa (1- \gamma) (B_1 - B_2)$ with $\gamma = \lambda /\kappa$ the ratio of the Land\'e factors (Appendix B).  We  obtain
\begin{align}
\eta= \frac{\gamma(B_1-B_2)}{B_1-B_2+\gamma B_2}  \leq 1 - \frac{B_{2}}{B_{1}} = \eta_\text{max}.
\label{eq:efficiency_dissip}
\end{align}
Its maximum value $\eta_\text{max}$, reached  in the absence of irreversible losses ($\gamma=1$), is determined by the ratio of the two magnetic fields. 
We evaluate the efficiency \eqref{eq:efficiency_dissip} using experimental data for different cycle durations, $\tau_\text{cycle} = \tau_\text{H}+\tau_\text{C} + 2 \tau$, by varying the heating and cooling times (Fig.~\ref{fig:Timeevolution}(a)). We find a constant value, i.e.~independent of the number of spin-exchange collisions, of  $\eta = 0.478 \pm 0.002$. We emphasize that the internal efficiency of the quantum Otto engine, $\eta_\text{int}= 1- |Q_\text{C}|/Q_\text{H} = 0.917 \pm 0.009$ (Appendix B) is close to the maximal value $\eta_\text{max} = 0.908$. We may therefore conclude that irreversible losses mainly occur during heat transfer processes, while the engine itself runs reversibly. The quantum heat engine is hence endoreversible. We further note that, since heat losses are  determined by the value of the Land\'e factors, they can in principle be reduced  by choosing different atomic species.

Second, we consider the average power of the quantum heat engine which reads
\begin{equation}
\label{eq:power}
P = \frac{Q_\text{H}-|Q_\text{C}|}{\tau_\text{cycle}} \leq \frac{Q_\text{H}}{\tau_\text{cycle}} \left( 1 - \frac{B_{2}}{B_{1}} \right).
\end{equation} 
We use the heat counting statistics to track its time evolution in  Fig.~\ref{fig:Timeevolution}(b). We observe that the  power (blue dots) increases with the number of inelastic collisions and reaches a maximum, $P_\mathrm{max}/k_\text{B}=30$ nK/ms, for a cycle time of  $960\,$ms. 
The corresponding number of inelastic collisions responsible for the heat exchange is almost twelve collisions total (6 spin-exchange collisions for the heating process and 6 for the cooling). 
This maximum nearly coincides with full population inversion between these two processes ($\ket{m_{F,\text{Cs}}=3} \leftrightarrow \ket{m_{F,\text{Cs}}=-3}$), in analogy to a laser. 
Good agreement with a theoretical model (red solid line) is observed (Appendix C). 
From a collisional perspective, the energy transfer with the atomic bath is optimal in the sense that it exchanges the maximum energy of six quanta, which can be stored in the machine, in exactly six spin-exchange collisions as a consequence of quantum engineering of the machine's and bath's spin states. The value of $P_\text{max}$ may be further optimized by enhancing the magnetic field difference, as well as the  collision rate and the collision cross-section by controlling the temperature or density of the Rb gas.
   
We finally  investigate the stability of the quantum Otto engine by analyzing the  relative power fluctuations via the Fano factor, which quantifies the deviation from a Poisson distribution \cite{fan47},
\begin{equation}
F_P = \frac{\sigma_{P}^2}{P} = \frac{\braket{P^{2}} - \braket{P}^{2}}{P},
\label{eq:Fano_factor}
\end{equation}
where $\sigma_{P}^{2}$ is the variance of the power, which we determine from the measured quasi-spin distributions (Appendix D). 
Figure~\ref{fig:Timeevolution}(c) displays the Fano factor as a function of the cycle time, with the absolute fluctuations $\sigma_P$ shown in the inset. We find super-Poissonian fluctuations ($F_P>1$) for short cycle times, indicating that the quantum engine is unstable in this regime, with  large relative power fluctuations. However, with increasing cycle time, the power increases faster than its variance, leading to a decrease in relative fluctuations. The transition to a Poissonian statistics ($F_P=1$) (red dashed line), with strongly reduced power fluctuations and significantly increased stability,  is located approximately at maximum power. This behavior follows from the finite Hilbert space  of the Cs machine and the saturation effect due to  the existence of an upper energy level. Importantly, the latter effect causes even the absolute value of the power fluctuations to decrease after on average six collisions  (Fig.~\ref{fig:Timeevolution}(c) inset). Power fluctuations could, in principle, also become sub-Poissonian ($F_P<1$), but this regime  is not seen experimentally due to experimental imperfections.

In conclusion, we have realized an endoreversible quantum Otto cycle using single Cs atoms interacting with a Rb bath. The key asset of this  machine is the quantum control over both the  few-level engine and  the atomic reservoir. This unique feature  allows us not only  to regulate and monitor the heat exchange between system and environment at the single-quantum level, but also to operate the quantum engine in a regime of  high efficiency, large power output and small power output fluctuations. The produced work could in principle  be extracted by coupling to the magnetic moment of the Cs atoms in a changing external magnetic field.  Our system provides a versatile experimental platform to elucidate fundamental new effects generated by quantum reservoir engineering, such as nonequilibrium atomic baths \cite{aba14,Alicki2015} and squeezed baths \cite{aba14a,Klaers2017},  as well  non-Markovian heat reservoirs by reducing the size of the Rb cloud \cite{Thomas2018,Pezzutto2019}.

We thank E.~Tiemann for providing us with the scattering cross-sections underlying our numerical model, and T. Busch and J.~Anglin for helpful comments on the manuscript.
This work was funded by Deutsche Forschungsgemeinschaft via Sonderforschungsbereich (SFB) SFB/TRR185 (Project No.~277625399) and Forschergruppe FOR 2724.

%%%%%%%%%%%%%%%%%%%%%%%%%%%%%%%%%%%%%%%%%%%%%%%%%%%%%%%%%%%%
%\newpage
%\section*{}
\newpage

\section*{SUPPLEMENTARY INFORMATION}

\subsection{A: Experimental procedures}
\label{sec:Experimental_procedure}

We start our experimental sequence by preparing an ultracold Rb gas in the magnetic field insensitive state $\Ket{F_{\text{Rb}} = 1, m_{F, \text{Rb}} = 0}$ and, at a distance of $\approx 200$ $\si{\micro\metre}$, a small sample of laser cooled Cs atoms.  The Cs atoms are further cooled and optically pumped into the $\Ket{F_{\text{Cs}} = 3, m_{F, \text{Cs}} = 3}$ hyperfine ground state by employing degenerate Raman sideband-cooling \cite{Kerman2000}. A species-selective optical lattice \cite{Schmidt2016} transports the Cs atoms into the Rb cloud.
MW radiation prepares the bath atoms in the state $\Ket{F_{\text{Rb}} = 1, m_{F, \text{Rb}} = -1}$. The starting point of the heat engine cycle is defined by switching off the optical lattice potential. After a predefined time $t_i$, the Cs-Rb interaction is stopped by freezing the positions of the Cs atoms using the optical lattice, and pushing the Rb cloud out of the trap with a resonant laser pulse. 
State-selective fluorescence imaging of the Cs atoms completes the procedure \cite{Schmidt2018}.

The high-energy and low-energy baths are interchanged by transferring the Rb atoms from $\ket{F_\text{Rb}=1,m_{F,\text{Rb}}=-1}$ to $\ket{F_\text{Rb}=1,m_{F,\text{Rb}}=+1}$ and vice versa
using two successive Landau-Zener sweeps. 
The transfer takes $\sim 4.4$~\si{\milli\second}, which is fast enough to avoid spin-exchange interactions during the state change of the bath. 

\begin{figure}
\centering
\includegraphics[width=8cm]{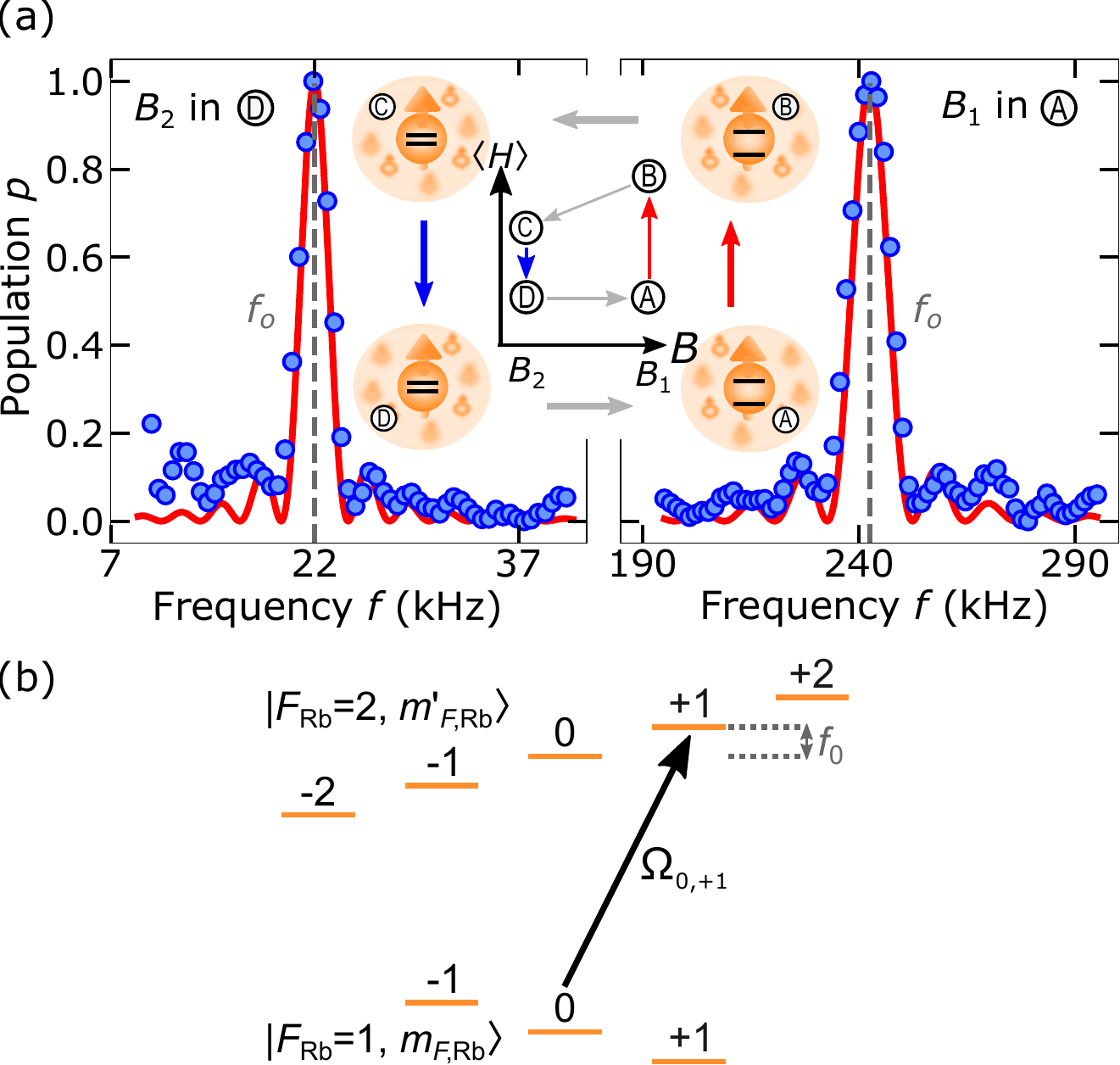}
\caption{\textbf{Magnetic field extraction.} Rb microwave spectra for extraction of the magnetic fields $B_1$ and $B_2$ (a) and corresponding transition scheme (b). Center of (a) illustrates the engine cycle and the corresponding Zeeman energy splitting of a Rb bath atom. Red lines correspond to the theory curves and blue dots are experimental data. These measurements yielding magnetic fields $B_{1} = 346.5 \pm 0.2 $ mG and $B_{2} = 31.6 \pm 0.1 $ mG. Measured spectra confirm similar magnetic fields for B and C.}
\label{fig:BfieldMeasurement}
\end{figure}
The two magnetic fields $B_1$ and $B_2$ defining the quantization axis for the engine operation, are measured using Rb microwave spectroscopy on the $\ket{F_\text{Rb}=1,m_{F,\text{Rb}}=0} \rightarrow \ket{F_\text{Rb}=2,m_{F,\text{Rb}}=+1}$ transition. 
The population of the Rb atoms in state $\Ket{F_\text{Rb}=2, m_{F,\text{Rb}}= +1}$ is detected by standard absorption imaging, using a time-of-flight measurement (Fig. \ref{fig:BfieldMeasurement}).
We fit the measured data with a standard model to extract the transition frequency, which translates into a magnetic field value using the Breit-Rabi formula \cite{HakenWolf}. We find typical errors of the order of $0.1\,$mG.

The magnetic field changes extracting work of the engine have to be adiabatic, i.e., preserving the populations $p_{n}$. The adiabaticity condition writes $\dot{\omega_\text{lar}}/\omega_\text{lar}^2 \ll 1$, where $\omega_\text{lar}= |g_F^\text{Rb}| \mu_\text{B} B/ \hbar$ is the Larmor frequency. It can therefore  be expressed as 
\begin{equation}
A\equiv \frac{\hbar \dot{B}}{|g_{F}| \mu_\text{B} B^{2}} \ll 1.
\end{equation}
Experimentally, we linearly vary the magnetic field from $B_{1}= 346.5 \pm 0.2$ mG to $B_{2}= 31.6 \pm 0.1$ mG in  a time scale of 10 ms, yielding values of $A(B_1)=0.2 \times 10^{-3}$ and $A(B_2)=14 \times 10^{-3}$, thus fulfilling the adiabatic condition at any time during the variation of the magnetic field. Moreover, the time scale of the magnetic field variation is faster than the time scale associated to the spin exchange collisions (see number of collisions over time in Fig.~\ref{fig:Timeevolution}). Hence, the populations $p_{n}$ are constant during the isentropic processes ($\text{B} \rightarrow \text{C}$ and  $\text{D} \rightarrow \text{A}$).    \\

%%%%%%%%%%%%%%%%%%%%%%%%%%%%%%%%%%%%%%%%%%%%%%%%%%%%%%%%%%%%%%

\subsection{B: Efficiency of the endoreversibe machine}
\label{ssec:Efficiency}
We calculate the efficiency by  distinguishing two different forms of heat exchange. 
First, we consider the respective  energies given $(Q_1)$ and taken $(Q_2)$ by the baths,  where $Q_1-|Q_2|$ is the energy turnover of the reservoirs per cycle. Second, we consider the energies absorbed ($Q_\text{H}$) and rejected ($Q_\text{C}$) from the engine, where $Q_\text{H}-|Q_\text{C}|$ is the energy turnover of the machine. Both quantities differ because of the different atomic Landé factors of Cs and Rb. The difference $Q_\text{L}= \left(Q_1-|Q_2|\right)-\left(Q_\text{H}-|Q_\text{C}|\right)$ is  dissipated via elastic collisions and irreversibly lost. We macroscopically model it  as a heat leak from the high-energy reservoir. Using the population distribution of the quasi-spin levels at the cycle points in Fig.~\ref{fig:QuantumEngine}, the individual heats can be calculated, leading to 
\begin{align} 
Q_\text{L} = &\left( Q_1-|Q_2| \right)-\left(Q_\text{H}-|Q_\text{C}|\right) \nonumber \\
= &\left(\sum_{n}n  \left[p_{n}^\text{B}-p_{n}^\text{A} \right]\kappa B_{1}
-\left | \sum_{n}n \left[p_{n}^\text{D}-p_{n}^\text{C} \right]\kappa B_{2}\right |\right)  \nonumber \\
& -\left(\sum_{n}n \left[p_{n}^\text{B}-p_{n}^\text{A} \right]\lambda B_{1} -\left |\sum_{n}n \left[p_{n}^\text{D}-p_{n}^\text{C}\right] \lambda B_{2}\right | \right).
\end{align}
Due to preservation of populations during adiabatic strokes, we can further use $p_{n}^\text{D}=p_{n}^\text{A}$ and $p_{n}^\text{B}=p_{n}^\text{C}$, yielding the expression for the dissipated heat
\begin{align}
   Q_\mathrm{L} &= \sum_n n \left (p_n^B - p_n^A \right) (\kappa - \lambda) (B_1 - B_2).
\end{align}
The efficiency is calculated as the work, $|W|= Q_\text{H}-|Q_\text{C}|$, produced by the engine, divided by the energy provided by the high-energy bath, $Q_\text{H}+Q_\text{L}$. 
Using $p_{n}^\text{D}=p_{n}^\text{A}$, $ p_{n}^\text{B}=p_{n}^\text{C}$ and $\gamma=\lambda/\kappa$, we find 
\begin{equation}
\eta = \frac{Q_\text{H}-|Q_\text{C}|}{Q_\text{H}+Q_\text{L}}= \frac{\gamma(B_1-B_2)}{B_1-B_2+\gamma B_2}.
% =\frac{\lambda (B_1-B_2)}{\kappa(B_1-B_2+\frac{\lambda}{\kappa}B_2)}
\end{equation}
The internal efficiency of the engine is computed as the ratio of the produced work $|W|$  and the heat absorbed by the machine $Q_\text{H}$:
\begin{equation}
\eta_\mathrm{int} =   \frac{Q_\text{H}-|Q_\text{C}|}{Q_\text{H}}= 1-\frac{B_2}{B_1}.
\end{equation}
It corresponds to the efficiency without a  leak ($\gamma=1$).

%%%%%%%%%%%%%%%%%%%%%%%%%%%%%%%%%%%%%%%%%%%%%%%%%%%%%%%%%%%%%
\subsection{C: Microscopic model and number of collisions}
\label{ssec:NumbersOfCollisions}

The quantum heat exchange between engine and bath is based on the understanding of individual spin-exchange collisions. In general, the spin-collision rate $\Gamma^{m_F \to m_F \pm1}$ is different both for every initial state $m_F$ and for the direction, i.e., $\Delta m_F = \pm 1$. The individual rates are well known from coupled-channel calculations of the molecular interaction potential between Rb and Cs \cite{Schmidt2019}. 
These rates allow us to describe the evolution with a rate model \cite{Bouton2020} that captures the spin dynamics and yields excellent agreement with the experimental data. 
From these rates, we also compute the mean number of spin collisions $N_\mathrm{spin}$ within a cycle duration $t=t_\text{D}$ in two steps. First, we calculate the time-averaged collision rate as the sum of time-averaged collision rates during heating (exothermal spin collisions) and cooling (endothermal spin collisions) as
\begin{equation}
\begin{split}
\braket{\Gamma (t)} &= \braket{\Gamma_{\text{A} \rightarrow \text{B}} (t)} + \braket{\Gamma_{\text{C} \rightarrow \text{D}} (t)} \\
&= \sum_{m_F=+3}^{-2} p_{m_F}(t)\Gamma_{\text{A} \rightarrow \text{B}}^{m_F \rightarrow m_F -1} \\
& + \sum_{m_F=+2}^{-3} p_{m_F}(t)\Gamma_{\text{C} \rightarrow \text{D}}^{m_F \rightarrow m_F +1}
%&= \sum_{m_F=+2}^{-3} p_{m_F}(t)\Gamma_{\text{A} \rightarrow \text{B}}^{m_F \rightarrow m_F +1} \\
%&+ \sum_{m_F=+3}^{-2} p_{m_F}(t)\Gamma_{\text{C} \rightarrow \text{D}}^{m_F \rightarrow m_F -1}.
\end{split}
\label{eq:mean_collision_rate}
\end{equation}
Second, we integrate these rates during the heating and cooling to obtain the number of collisions within cycle time $t$ as
\begin{equation}
\begin{split}
N_{\text{spin}}(t) &= N_{\text{A} \rightarrow \text{B}} + N_{\text{C} \rightarrow \text{D}} \\
&= \int_{0}^{t_\text{B}} (\braket{\Gamma_{\text{A} \rightarrow \text{B}} (t^{\prime})}  dt^{\prime} +
\int_{t_\text{C}}^{t_\text{D}} \braket{\Gamma_{\text{C} \rightarrow \text{D}} (t^{\prime})}) dt^{\prime}.
\end{split}
\label{eq:number_collisions}
\end{equation}
In order to close the cycle, the inital and final Cs states before and after a cycle have to be the equal, leading to the condition $ N_{\text{A} \rightarrow \text{B}} = N_{\text{C} \rightarrow \text{D}}$.

%%%%%%%%%%%%%%%%%%%%%%%%%%%%%%%%%%%%%%%%%%%%%%%%%%%%%%%%%%%%%%
\subsection{D: Fluctuations of the quantum machine}
\label{ssec:EntropyAndFluctuations}
To extract the fluctuations of the engine, Eq.~(\ref{eq:Fano_factor}), we calculate the power, Eq.~(\ref{eq:power}), via $P=|W|/\tau_\text{cycle}$. The cycle time $\tau_\text{cycle}=t_\text{D}$ is experimentally controlled, and we assume that it is a fixed parameter not adding further fluctuations  to the power-output fluctuations. Therefore, we can restrict the calculation to the fluctuations $\sigma_W$ of work $W$ as $\sigma_W^2 = \langle W^2 \rangle - \langle W \rangle^2$. 
The work is given by the difference of energy absorbed by and rejected from the engine $|W|=Q_\text{H} - |Q_\text{C}|$, and hence 
\begin{align}
  \sigma_W^2 &= \sigma_{Q_\text{H}}^2 + \sigma_{Q_\text{C}}^2  \nonumber \\
  &= \langle Q_\text{H}^2 \rangle - \langle Q_\text{H} \rangle^2 + \langle Q_\text{C}^2 \rangle - \langle Q_\text{C} \rangle ^2. \label{eq:WorkFluctuations}
\end{align}
The averages and variances of heat absorbed or rejected depend on the energy differences at the different points during the cycle, for example, $Q_\text{H} = E(t_\text{B},B_1) - E_0(t_0,B_1)$. Here $E(t_i,B_j)= \sum_n p_n^i (t_i) \,n \lambda B_j$ can be computed from the measured quantum-level populations $\{p_n^i\}$ of level $n$ at point $i=\text{A,B,C,D}$ during the cycle and the magnetic field $B_j$ $(j=1,2)$, together with mean energy and variance. 
Then, the fluctuations $\sigma_Q^2 $ of heat $Q$ exchanged when changing the engine's probability distribution from point $i$ to point $f$ at a magnetic field $B_j$ reads
\begin{align}
  \sigma_Q^2  =& \sum_n \left (p_n^f(t_f) + p_n^i(t_i) \right )  (n\lambda B_j)^2  \nonumber \\
  & - \left\{ \left [ \sum_n p_n^f(t_f) n \lambda B_j \right ]^2 + \left [ \sum_n p_n^i(t_i) n \lambda B_j \right ]^2  \right \},
\end{align}
where, using the notation of Fig.~1(c), for $Q_\text{H}$ $i=0$, $f=\text{B}$, and $B_j=B_1$, and for $Q_\text{C}$ $i=\text{C}$, $f=\text{D}$, and $B_j=B_2$. 
Inserting these expressions into Eq.~(\ref{eq:WorkFluctuations}) allows us to compute the work fluctuations for every cycle time $\tau_\text{cycle} =t_\text{A}$ and thereby the variance of the output power fluctuations $\sigma_P^2$. 

\end{document}